\documentclass[prb,preprint]{revtex4-1} 

\usepackage{graphicx}
\usepackage{dcolumn}
\usepackage{bm}

\begin{document}

\title{Inverted approach to the inverse scattering problem: complete solution
of the Marchenko equation for a model system}
\author{Matti Selg}
\affiliation{Institute of Physics of the University of Tartu, Ravila 14c,
50411, Tartu, Estonia}
\date{\today}

\begin{abstract}
An example of full solution of the inverse scattering problem on the half
line (from 0 to $\infty $) is presented. For this purpose, a simple
analytically solvable model system (Morse potential) is used, which is
expected to be a reasonable approximation to a real potential. First one
calculates all spectral characteristics for the fixed model system. This way
one gets all the necessary input data (otherwise unobtainable) to implement
powerful methods of the inverse scattering theory. In this paper, the
multi-step procedure to solve the Marchenko integral equation is described
in full details. Excellent performance of the method is demonstrated and its
combination with the Marchenko differential equation is discussed. In
addition to the main results, several important analytic properties of the
Morse potential are unveiled. For example, a simple analytic algorithm to
calculate the phase shift is derived.

\end{abstract}

\maketitle

\section{Introduction}

Strict mathematical criteria for the unique solution of the one-dimensional
inverse scattering problem have been formulated long ago, thanks to the
outstanding contributions by many researchers \cite{Levinson1, Levinson2,
Bargmann1, Bargmann2, GL, Jost1, Jost2, Marchenko1, Marchenko2, Krein1,
Krein2} (see \cite{Chadan} for a historical overview). However, in spite of
the perfect consistency and mathematical beauty of the concept, these
rigorous criteria have rarely been used in practice, because the necessary
input data are very difficult to obtain. Indeed, one has to know the full
energy spectrum of the bound states $E_{n}<0$ ($n=0,1,2,...,N-1$) and the
full energy dependence (from 0 to $\infty $) of the phase shift $\delta (E)$
for the scattering states ($E>0$). Even if the mentioned obligatory
requirements were fulfilled (which is never the case), this would only be
sufficient to construct an $N$-parameter family of phase-equivalent and
isospectral potentials. One has to somehow fix $N$ additional parameters,
the so-called norming constants, in order to ascertain the potential
uniquely.
\begin{figure}[tbh]
\includegraphics[width=0.8\textwidth]{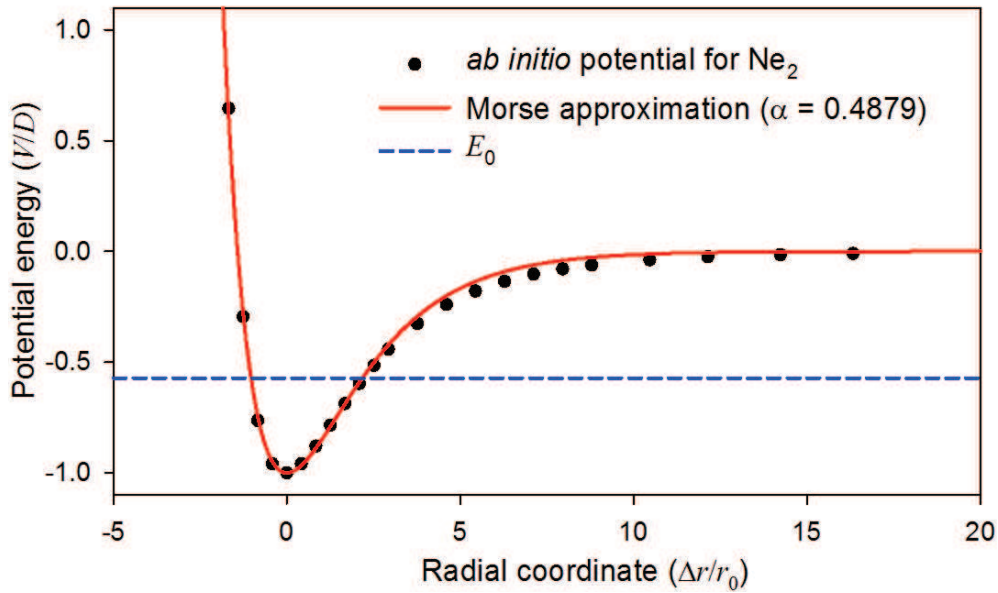}
\caption{\label{fig:Ne2}
An illustration to the physical background of the proposed approach.}
\end{figure}
On the other hand, the shape of a real potential is not arbitrary, but often
it looks like the curve in Fig. \ref{fig:Ne2}. In this figure, a comparison
is made between an \textit{ab initio} potential for Ne$_{2}$ molecule \cite%
{NePot} and its Morse approximant, calculated according to the formula \cite%
{Morse29}%
\begin{equation}
\frac{V(r)}{D}=\exp \left[ -\frac{2\alpha \left( r-R_{e}\right) }{r_{0}}%
\right] -2\exp \left[ -\frac{\alpha \left( r-R_{e}\right) }{r_{0}}\right] ,
\label{F1}
\end{equation}%
taking $D=42.153$ K \cite{NePot, Merkt} and $R_{e}=3.0895$ \AA\ \cite{NePot}%
. The Morse curve in Fig. \ref{fig:Ne2} was calculated with $r_{0}\equiv 
\sqrt{\hbar ^{2}/(2mD)}=0.2388$ \AA ,\ $\alpha =0.4879/r_{0},$ and $\Delta
r\equiv r-R_{e}$. It has only one bound state: $E_{0}/D=-0.5716,$ and
practically the same value has been determined for the single discrete level
of the Ne$_{2}$ molecule, both theoretically \cite{NePot} and experimentally 
\cite{Merkt}.

Fig. \ref{fig:Ne2} was presented for illustrative purposes, to continue the
above discussion. Namely, in principle, the unknown norming constants for a
real potential can be treated as variational parameters. One may try to
guess their correct values by fitting to Eq. (\ref{F1}). Moreover, one might
even think about using this simple model potential to examine different
solution schemes of the inverse scattering theory, and this is the main idea
developed in this paper. Morse potential has many unique analytic
properties, which makes it especially suitable for modelling. In particular,
for this model system one can ascertain all the necessary spectral
characteristics to any desired accuracy. Thereafter, one can use them as
input data, instead of getting these data from real experiments.

The idea to invert an inverse problem is, of course, tautological but not at
all meaningless. Indeed, despite the immense computational power which is
now available to the researchers, one hardly could find any example of
complete solution of the inverse scattering problem on the half line. The
main reason for this deficiency is the lack of the necessary input data, and
consequently, there seems to be no motivation to perform such studies. Then,
what is the motivation for the present study? First of all, in author's
opinion there is a debt of honor to be paid to the researchers who developed
this beautiful theory. Nowadays people tend to consider the inverse scattering
a 'pedagogical' problem which is of little scientific interest. Unfortunately,
such an ignorant view is not based on the real knowledge, but is a mere belief.
Implementation of the methods of the inverse scattering theory is not at all
a trivial task. On the contrary, this is a complex and computationally
demanding multi-step procedure which has to be performed with utmost accuracy.
In this paper, we are going to describe this procedure for a very simple model
system, to make it understandable for a wide audience of potential readers.
This can be considered a first step towards the real implementation of the
method. Another, more specific motive for this work is the obvious need to
clarify some conceptual details, as will be explained in Sec. II.

Performing an applicability test of the Marchenko method is not the only
issue addressed in this paper. In author's opinion, the model system itself
deserves special attention. Useful properties of the Morse potential, as
well as the solution to the related Schr\"{o}dinger equation already given
by Philip Morse himself \cite{Morse29}, are well known to physicists. Since
the potential is shape-invariant, this solution can even be found by pure
algebraic means, using the Gendenshtein's recipe \cite{Gendenshtein}.
However, in this context we are talking about the physically correct linear
combination of the two special solutions to the Schr\"{o}dinger equation.
There are lots of systems which can be analyzed in exactly the same way (the
simplest one is the harmonic oscillator), but within this pattern the
feature which makes Morse potential really unique is overlooked. Namely, for
a Morse potential the two linearly independent solutions to the Schr\"{o}%
dinger equation can always be easily ascertained analytically (even if $D<0$%
!). This is not the case for a harmonic oscillator or any other popular
model system.

Consequently, the approach based on Eq. (\ref{F1}) is not just an option
among many others, but it is often the most appropriate choice for modelling
a real quantum system. Moreover, one may construct a more reliable model
consisting of several Morse-type components, still preserving the exact
solubility of the Schr\"{o}dinger equation. Such technically rather complex
developments are not analyzed in this paper (see, e.g., \cite{Selg2001} and 
\cite{SelgJCP} for more details), because here we concentrate on solving
the inverse not the direct Schr\"{o}dinger problem. For this reason, the
model has been chosen to be as simple as possible.

Nevertheless, as a matter of fact, the scattering properties of the Morse
potential have not found much attention so far, and there still is some work
to do. In Sec. III, a simple analytic formula for calculating the phase
shift is derived. Unexpectedly, in addition to this specific result, there is
a pure mathematical outcome which is directly related to Eq. (\ref{F1}).
Namely, a very accurate algorithm for evaluating the Riemann-Siegel function
(see \cite{Edwards}) was obtained, as a kind of bonus for calculating the
phase shift.

The remaining parts of the paper are organized as follows. The details of
calculating the kernel of the Marchenko equation are described in Sec. IV
and the solution method itself is under examination in Sec. V. Finally, Sec.
VI concludes the work.

\section{Marchenko integral equation}

The basis for this paper is the integral equation%
\begin{equation}
A(r,t)=A_{0}(r+t)+\int\limits_{r}^{\infty }A(r,s)A_{0}(s+t)ds,\ t\geq r,
\label{F2}
\end{equation}%
derived by Marchenko \cite{Marchenko1}. The Marchenko method is preferred,
because the kernel of Eq. (\ref{F2}) is directly related to the main
spectral characteristics of the system, while an alternative approach, the
Gelfand-Levitan method \cite{GL}, requires an additional calculation of the
Jost function.

The kernel of Eq. (\ref{F2}) reads%
\begin{eqnarray}
A_{0}(x)=\frac{1}{2\pi }\int\limits_{-\infty }^{\infty }\left[ S(k)-1\right]
\exp (ikx)dk-\sum_{n}s_{n}^{2}\exp (-\gamma _{n}x),  \label{F3} \\
k\equiv \sqrt{E/C},\ \gamma _{n}\equiv \sqrt{-E_{n}/C},\ S(k)\equiv \exp %
\left[ 2i\delta (k)\right] ,  \nonumber
\end{eqnarray}%
where $C\equiv \hbar ^{2}/(2m)$ and $s_{n}$ are the norming constants for
the Jost solution of the related Schr\"{o}dinger equation:%
\begin{equation}
f(i\gamma _{n},r)=\exp (-\gamma _{n}r)+\int\limits_{r}^{\infty }A(r,x)\exp
(-\gamma _{n}x)dx,  \label{F4}
\end{equation}%
so that $f(i\gamma _{n},r)\rightarrow \exp (-\gamma _{n}r)$ as $r\rightarrow
\infty $, and%
\begin{equation}
s_{n}^{2}\int\limits_{0}^{\infty }\left[ f(i\gamma _{n},r)\right] ^{2}dr=1.
\label{F5}
\end{equation}%
If one is able to solve Eq. (\ref{F2}) then the potential can be easily
determined:%
\begin{equation}
V(r)=-2C\frac{dA(r,r)}{dr}.  \label{F6}
\end{equation}

The key formulas were given in full detail, in order to clarify the issue
mentioned above. Namely, in several highly appreciated overviews either the
Marchenko equation itself or its kernel or both of them are given
incorrectly (see Refs.~\onlinecite{Chadan1}, p. 73; \onlinecite{Chadan}, p. 79;
\onlinecite{Chadan2}, p. 732; \onlinecite{Chadan3}, p. 178). The dispute
concerns the sign of the second term in Eq. (\ref{F3}), which is 'plus' in
all these overviews. According to Marchenko's original works (see, e.g.,
Refs.~\onlinecite{AgrMarch}, p. 62; \onlinecite{Marchenko}, p. 218), on the
contrary, there should be 'minus' as in Eq. (\ref{F3}), and this is
definitely true. Indeed, the original Marchenko equation can be rewritten as%
\begin{eqnarray*}
A(r,t) &=&-B_{0}(r+t)-\int\limits_{r}^{\infty }A(r,s)B_{0}(s+t)ds, \\
B_{0}(x) &=&\frac{1}{2\pi }\int\limits_{-\infty }^{\infty }\left[ 1-S(k)%
\right] \exp (ikx)dk+\sum_{n}s_{n}^{2}\exp (-\gamma _{n}x).
\end{eqnarray*}%
Consequently, taking $A_{0}(x)\equiv -B_{0}(x)$, one comes to Eqs. (\ref{F2}%
)-(\ref{F3}). Most likely, the wrong sign given in Refs.~\onlinecite{Chadan}%
, \onlinecite{Chadan1}-\onlinecite{Chadan3} is just a human error. However, it
is misleading (especially in the pedagogical context) and should be corrected.
On the other hand, it is quite surprising that this error has not even been
noticed until now. Indirectly, it means that no one has attempted to put the
Marchenko method into practice, which provides another motivation for the
present study.

Let us specify the problem and the scheme for its solution. First, one
calculates the spectral characteristics for a simple model potential
described by Eq. (\ref{F1}). Then one determines the kernel of the Marchenko
equation according to Eq. (\ref{F3}). Finally, one solves Eq. (\ref{F2}) and
calculates the related potential according to Eq. (\ref{F6}). If the
procedure is successful then the initial potential is recovered. Throughout
this paper, dimensionless units for the energy and radial coordinate will be
used, taking $r_{0}\equiv \sqrt{\hbar ^{2}/(2mD)}=1,$ $D=1,$ and
consequently, $C=1.$

For a Morse potential, the norming constants $s_{n}$ can be easily
ascertained analytically. Let us fix, for example, $\alpha =2/3.$ Then the
system has only one bound state:%
\begin{equation}
E_{n}=-D(1-\frac{n+1/2}{a})^{2}\rightarrow E_{0}=-D(1-\frac{1}{2a})^{2},
\label{F7}
\end{equation}%
where $a\equiv \sqrt{D/C}/\alpha .$ Consequently, in our unit system, $%
a=1/\alpha $ and $\gamma _{0}=2/3=\alpha .$ Therefore, this particular case
is especially suitable for illustrative purposes and only this case will be
examined further on. (The same approach, with minor adjustments concerning
the different values for $\alpha $ and $\gamma _{0}$, can be applied to the
model potential shown in Fig. \ref{fig:Ne2}). To conclusively fix the model,
let us assume that the minimum of the potential is located at $%
R_{e}=2.5r_{0} $.

At first sight it may seem that the system has another level exactly at the
zero point. Indeed, if one takes $\alpha =2/3$ then $a=3/2$ and, according
to Eq. (\ref{F7}), $E_{1}=0$. Actually, however, this is not the case,
because the coordinate here ranges from 0 to $\infty $, not from $-\infty $
to $\infty $ as for a classical Morse oscillator. Therefore, as can be
easily proved, the possible additional level is shifted slightly above (not
below) the dissociation limit, becoming a scattering state. Another nuance
is that, strictly speaking, one should form a linear combination of both
solutions of the Schr\"{o}dinger equation, to fix the correct regular
solution. Again, it can be shown (see, e.g., \cite{Selg2006}) that the
second special solution is of next to no importance, so that the physical
solution reads%
\begin{equation}
\Psi _{0}=N_{0}\exp (-y/2)y  \label{F8}
\end{equation}%
as in the 'normal' Morse case. Here $N_{0}$ is the related norming constant
and a new dimensionless coordinate $y\equiv 2a\exp \left[ -\alpha \left(
r-R_{e}\right) \right] $ was introduced.

The parameter $N_{0}$ as well as the norming constant for the Jost solution (%
\ref{F4}) can be easily determined:%
\begin{equation}
s_{0}^{2}=\frac{y_{0}^{2}\alpha }{1-\int\limits_{y_{0}}^{\infty }\exp (-y)ydy%
},\ y_{0}\equiv 2a\exp \left( \alpha R_{e}\right) .  \label{F9}
\end{equation}%
The second term in the denominator is negligible, so that, to a high
accuracy, $s_{0}^{2}=y_{0}^{2}\alpha =6\exp (10/3)$. This is the value which
we are trying to recover. However, to clarify the sign problem described above,
the quantity $s_{0}^{2}$ will be treated as a variational (not definitely
positive!) parameter. We will see that its expected value is indeed the correct
value.

\section{Phase shift for the Morse potential}

Let us see how to determine the full energy dependence of the phase shift,
subjected to a strong constraint%
\begin{equation}
\delta (0)-\delta (\infty )=N\pi   \label{F10}
\end{equation}%
($N=1$ in our case), according to the Levinson theorem \cite{Levinson1,
Levinson2}. The analytic background of the problem has been described
elsewhere \cite{SelgJCP} (see Eqs. (23) to (29) there). There are several
formally equivalent approaches to building the regular solution of the Schr%
\"{o}dinger equation for a scattering state. For example, we may proceed
from Eq. (25) of Ref.~\onlinecite{SelgJCP}. Then, after some analytical work,
we get the following special solutions:%
\begin{eqnarray}
\Psi _{1}(k,r) &=&\left\vert S\left( a,i\beta ;y\right) \right\vert \cos
\left\{ \alpha _{0}+\arg \left[ S\left( a,i\beta ;y\right) \right]
-kr\right\} ,  \label{F101} \\
\Psi _{2}(k,r) &=&\left\vert S\left( a,i\beta ;y\right) \right\vert \sin
\left\{ \alpha _{0}+\arg \left[ S\left( a,i\beta ;y\right) \right]
-kr\right\} ,  \nonumber
\end{eqnarray}%
with%
\begin{eqnarray}
S(a,i\beta ;y) &=&\sum_{n=0}^{\infty }B_{n},B_{0}=1,B_{1}=-\frac{ay}{2i\beta
+1},\ \beta \equiv k/\alpha ,  \nonumber \\
\ B_{n} &=&\frac{y}{n(2i\beta +n)}(-aB_{n-1}+\frac{y}{4}B_{n-2}),\ n=2,3,...,
\label{F12}
\end{eqnarray}
\begin{equation}
\alpha _{0}\equiv \beta \ln (2a)+\beta _{N}-B+kR_{e},  \label{F100}
\end{equation}%
$\beta _{N}\equiv \arg \left[ \left( i\beta +1\right) \left( i\beta
+2\right) ...\left( i\beta +N\right) \right] $, and%
\begin{equation}
B\equiv \arg \left[ \frac{\Gamma (2i\beta )}{\Gamma (i\beta )}\right] ,
\label{F102}
\end{equation}%
($N=1$ in our case), $\Gamma $ being the gamma function.
\begin{figure}[tbh]
\includegraphics[width=0.8\textwidth]{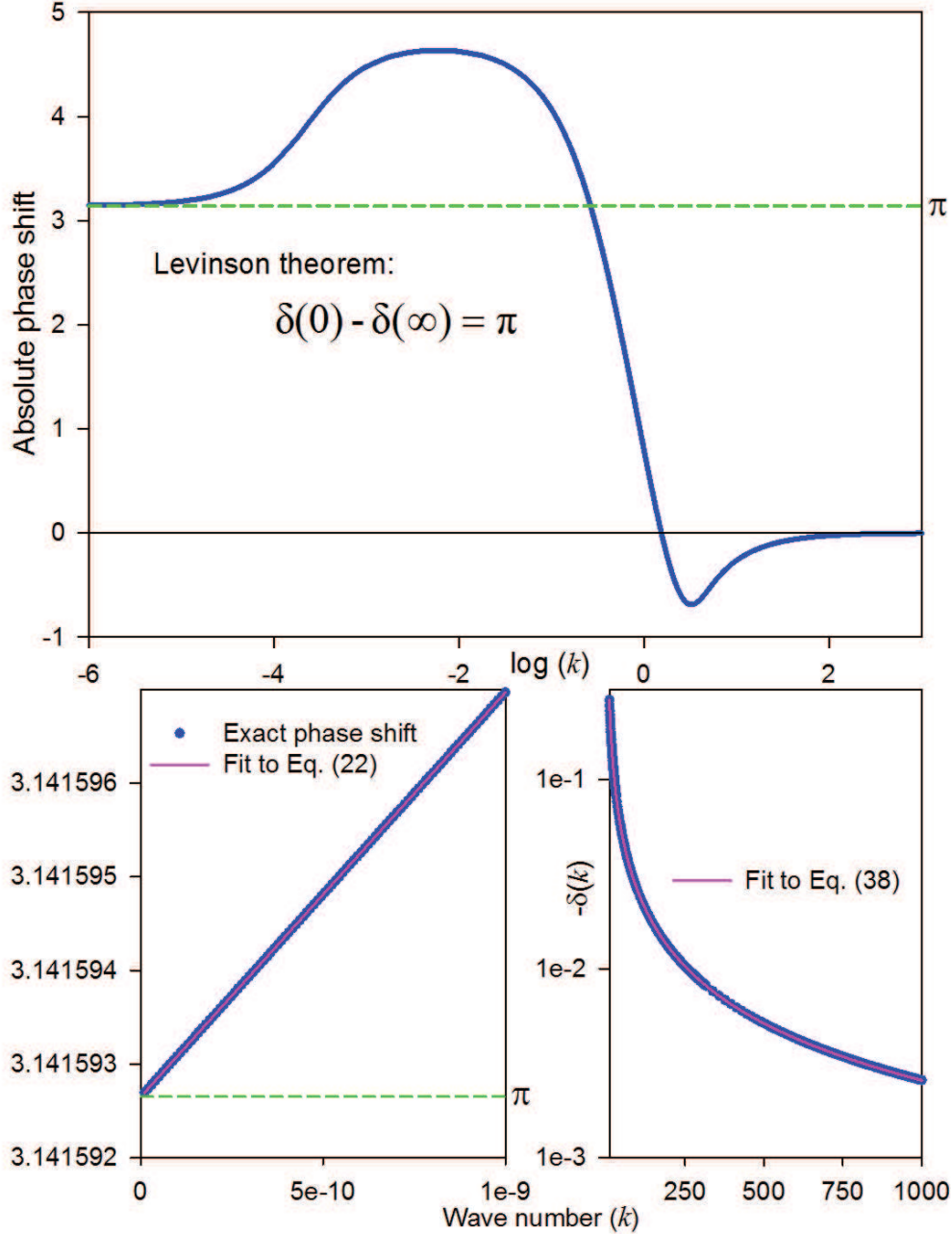} 
\caption{\label{fig:Phase}Phase shift for the potential Eq. (\ref{F1})
depending on the wave number (parameters are specified in Sec. II). Lower
graphs demonstrate the asymptotic behavior of $\delta (k)$.}
\end{figure}

\begin{figure}[tbh]
\includegraphics[width=0.8\textwidth]{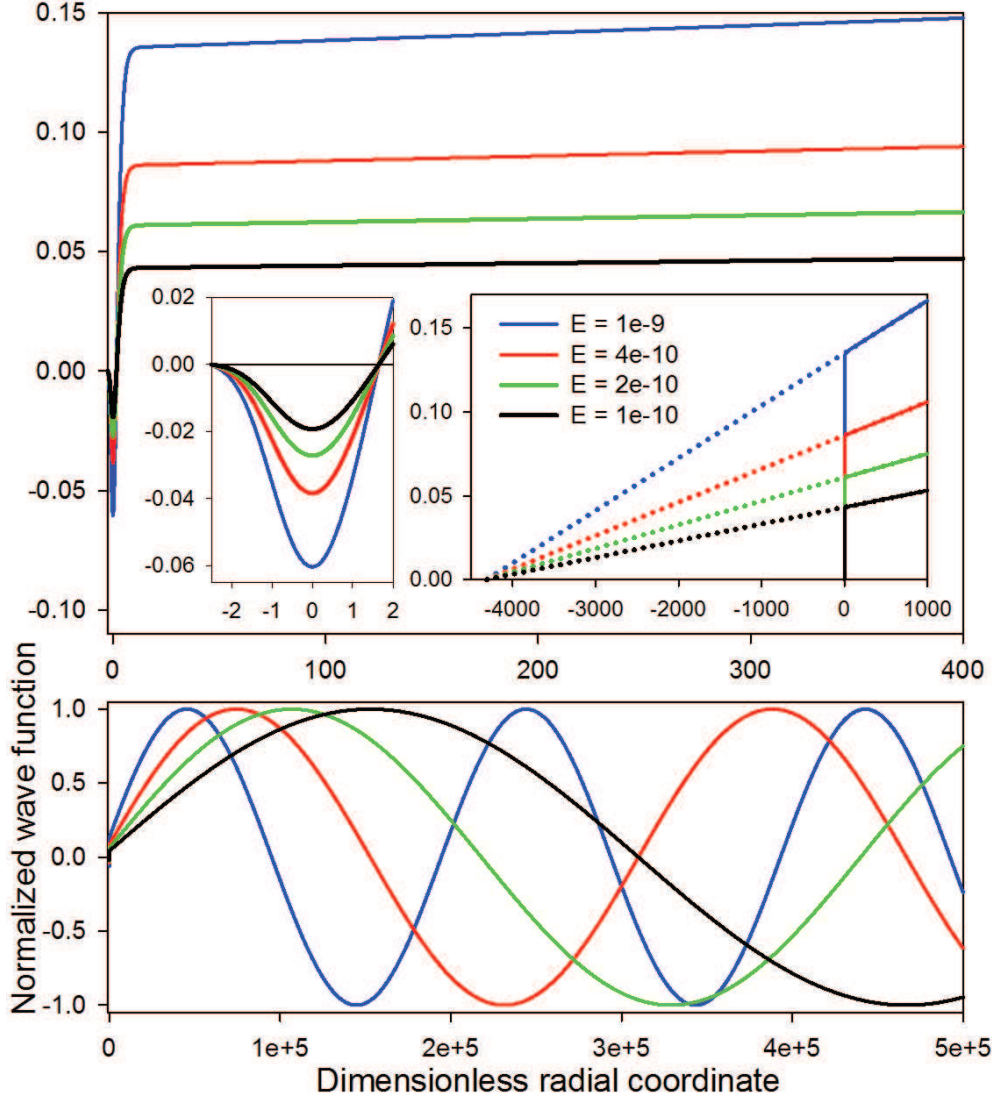} 
\caption{\label{fig:Wfunc}Wave functions for the scattering states at extremely
small energies. The same set of wave functions is shown in all graphs and
their insets, but note that the coordinate ranges are essentially different.}
\end{figure}

From the regularity requirement $\Psi (0)=A_{1}\Psi _{1}(0)+A_{2}\Psi
_{2}(0)=0,$ one gets the ratio of the coefficients%
\begin{equation}
\frac{A_{1}}{A_{2}}=-\tan \left[ \alpha _{0}+\arg \left( S_{0}\right) \right]
,\ S_{0}\equiv S\left( a,i\beta ;y_{0}\right) .  \label{F103}
\end{equation}%
On the other hand, since $S(a,i\beta ;y)\rightarrow 1$ as $r\rightarrow
\infty $ \cite{SelgJCP}, one gets a formula for the phase shift:%
\begin{equation}
\tan \left[ \delta (k)\right] =\frac{\frac{A_{1}}{A_{2}}\cos \alpha
_{0}+\sin \alpha _{0}}{\frac{A_{1}}{A_{2}}\sin \alpha _{0}-\cos \alpha _{0}},
\label{F104}
\end{equation}%
and consequently, a general expression for the wave function $\Psi
=A_{1}\Psi _{1}+A_{2}\Psi _{2}$ (for simplicity we drop the arguments of the 
$S$-functions)%
\[
\Psi (k,r)=-\frac{A_{2}\left\vert S\right\vert }{\cos \left[ \alpha
_{0}+\arg \left( S_{0}\right) \right] }\sin \left[ kr+\arg \left(
S_{0}\right) -\arg \left( S\right) \right] .
\]%
The last formula is valid for any $r$, and also in the limit $r\rightarrow
\infty $. From this one immediately concludes that $A_{2}=-\cos \left[
\alpha _{0}+\arg \left( S_{0}\right) \right] $ and $A_{1}=\sin \left[ \alpha
_{0}+\arg \left( S_{0}\right) \right] ,$ which means that%
\begin{equation}
\Psi (k,r)=\left\vert S\right\vert \sin \left\{ kr+\arg \left( S_{0}\right)
-\arg \left( S\right) \right\} ,  \label{F11}
\end{equation}%
and most importantly,%
\begin{equation}
\delta (k)=\arg \left( S_{0}\right) .  \label{F13}
\end{equation}

This is the simple formula mentioned in Sec. I. Combined with Eq. (\ref{F12}%
), it enables to easily ascertain the phase shift for any scattering state
and thus, its full dependence on energy (or wave number). The result for
the specified Morse potential is shown in Fig. \ref{fig:Phase} (main graph).
Note that the \textit{k}-scale there is logarithmic, so that the figure
involves 18 orders of magnitude in the energy space. Full agreement with the
Levinson theorem (\ref{F10}) can be seen as well.

Fig. \ref{fig:Wfunc} demonstrates the behavior of the scattering wave
functions as $E\rightarrow 0.$ In general, the wave functions can be
calculated using Eqs. (\ref{F12}), (\ref{F11}) and (\ref{F13}). Note,
however, that the $S$-functions are defined as \cite{Selg2001} 
\begin{equation}
S(a,c;x)\equiv \exp (-x/2)\Phi (-a+c+1/2,2c+1;x),  \label{S1}
\end{equation}%
where%
\begin{equation}
\Phi (a,c;x)=1+\frac{ax}{1!c}+\frac{a(a+1)x^{2}}{2!c(c+1)}+....  \label{S2}
\end{equation}%
Therefore, if $E\rightarrow 0$ and consequently, $i\beta \rightarrow 0$,
then it is more appropriate to explicitly use Eq. (\ref{S2}). As a result,
one gets the following rapidly converging expression:%
\begin{equation}
S(a,i\beta ;y)=\exp (-y/2)\left\{ 1-y+i\beta y\left[ 3-\frac{0!}{\left(
2!\right) ^{2}}y-\frac{1!}{\left( 3!\right) ^{2}}y^{2}-\frac{2!}{\left(
4!\right) ^{2}}y^{3}-...\right] \right\} ,  \nonumber
\end{equation}%
which was actually used to compose Fig. \ref{fig:Wfunc}.

As the energy is extremely low, linear coordinate dependence can be seen in
a wide region, which is in full agreement with the well-known rule%
\begin{equation}
\delta (k)=N\pi -\arctan (ka_{0}),  \label{S4}
\end{equation}%
where $a_{0}$ is the scattering length. From Eqs. (\ref{F11}), (\ref{F13})
and (\ref{S4}) it follows that%
\[
\Psi (k,r)\rightarrow \sin \left[ kr+\delta (k)\right] \rightarrow \sin
\left( kr\right) +\tan \left[ \delta (k)\right] \rightarrow k\left(
r-a_{0}\right) ,
\]%
and this is indeed the case, while $a_{0}=-4312.06224.$

Practically the same scattering length is obtained from the real \textit{k}%
-dependence of the phase shift (see the lower left graph in Fig. \ref%
{fig:Phase}), using Eqs. (\ref{F12}), (\ref{F13}) and (\ref{S1})-(\ref{S4}).

\subsection{An algorithm for the Riemann-Siegel function}

Usually, to calculate $\delta (k)$ there is no need to make any use of Eqs. (%
\ref{F100})-(\ref{F102}). In some special cases, however, the convergence of
the series (\ref{F12}) may be slow. Then it may be appropriate to combine Eqs.
(25) and (28) from Ref.~\onlinecite{SelgJCP}. As a result, again after some
analytical work, one comes to the following expression:%
\begin{equation}
\tan \left[ \alpha _{0}+\arg \left( S\right) \right] =\frac{\alpha
_{N}^{2}\beta }{y^{2N+1}}\exp (y)F\left( N,i\beta ;y\right) ,  \label{F131}
\end{equation}%
where $\alpha _{N}\equiv \left\vert \left( i\beta +1\right) \left( i\beta
+2\right) ...\left( i\beta +N\right) \right\vert ,$%
\begin{equation}
F\left( N,i\beta ;y\right) \equiv \frac{1-\frac{N^{2}+\beta ^{2}}{1!y}+\frac{%
\left( N^{2}+\beta ^{2}\right) \left[ \left( N-1\right) ^{2}+\beta ^{2}%
\right] }{2!y^{2}}+...}{1+\frac{\left( N+1\right) ^{2}+\beta ^{2}}{1!y}+%
\frac{\left[ \left( N+1\right) ^{2}+\beta ^{2}\right] \left[ \left(
N+2\right) ^{2}+\beta ^{2}\right] }{2!y^{2}}+...},  \label{F132}
\end{equation}%
and we assumed that $N=a-1/2$.

To exploit Eqs. (\ref{F131})-(\ref{F132}), one takes $y=y_{0}$, uses Eq. (%
\ref{F100}), and finally gets the phase shift according to Eq. (\ref{F13}).
The only problem is how to ascertain the parameter $B$ defined by Eq. (\ref%
{F102}). Fortunately, we can apply to the theory of gamma functions (see 
\cite{Bateman}, Sec. 1.2):%
\[
\Gamma (1/2+z)\Gamma (1/2-z)=\frac{\pi }{\cos \left( \pi z\right) },
\]

\begin{equation}
\Gamma (2z)=\frac{2^{2z-1}}{\sqrt{\pi }}\Gamma (z)\Gamma (1/2+z)=\frac{%
2^{2z-1}\sqrt{\pi }}{\cos \left( \pi z\right) }\cdot \frac{\Gamma (z)}{%
\Gamma (1/2-z)}.  \label{F133}
\end{equation}%
Taking here $2z=1/2+i\beta $, one easily gets the desired result:%
\begin{equation}
B=\arg \left[ \frac{\Gamma (2i\beta )}{\Gamma (i\beta )}\right] =\arg \left[
\Gamma (i\beta +1/2)\right] +2\beta \ln 2=\beta \ln \left( 8\pi \right)
+\arctan \left[ \tanh \left( \frac{\beta \pi }{2}\right) \right] +2\theta
(\beta ),  \label{F134}
\end{equation}%
where%
\begin{equation}
\theta (\beta )\equiv \arg \left[ \Gamma (1/4+i\beta /2)\right] -\beta /2\ln
\pi   \label{F135}
\end{equation}%
is the Riemann--Siegel theta function, whose behavior is well known as $%
\beta \rightarrow 0$ or $\beta \rightarrow \infty .$ Namely \cite{Edwards},%
\begin{equation}
\beta \rightarrow 0:\theta (\beta )=-\frac{\beta }{4}\left[ 2\gamma +\pi
+2\ln \left( 8\pi \right) \right] +\frac{\beta ^{3}}{24}\left[ \pi
^{3}+28\zeta (3)\right] -\beta ^{5}\left[ \frac{\pi ^{5}}{96}+\frac{31\zeta
(5)}{10}\right] +...  \label{F136}
\end{equation}%
and%
\begin{equation}
\beta \rightarrow \infty :\theta (\beta )=-\frac{\beta }{2}\ln \left( \frac{%
2\pi }{\beta }\right) -\frac{\beta }{2}-\frac{\pi }{8}+\frac{1}{48\beta }+%
\frac{7}{5760\beta ^{3}}+\frac{31}{80640\beta ^{5}}+...  \label{F137}
\end{equation}

Here $\gamma =0.5772156649...$ is the Euler-Mascheroni constant and $\zeta
(3)=1.2020569032...,\zeta (5)=1.0369277551...$ are the corresponding values
of the Riemann zeta function. In practice, Eq. (\ref{F136}) can be used if $%
\beta <0.1,$ while Eq. (\ref{F137}) is recommended for $\beta >20$. For the
intermediate region $\beta \in \left[ 0.1,20\right] $ an innovation can be
introduced. Namely, combining Eq. (\ref{F134}) with a useful formula for
calculating $\arg \left[ \Gamma (i\beta +1/2)\right] $ \cite{Chebotarev}, we
get%
\begin{equation}
2\theta (\beta )=\beta \left[ \ln \left( \frac{\sqrt{(1+4\beta ^{2})}}{4\pi }%
\right) -1\right] -\arctan \left[ \tanh \left( \frac{\beta \pi }{2}\right) %
\right] -\frac{I(\beta )}{2},  \label{F138}
\end{equation}%
where \cite{Selg2001}%
\begin{equation}
I(\beta )\equiv \int\limits_{0}^{\infty }\left( \coth t-\frac{1}{t}\right)
e^{-t}\sin (2\beta t)\frac{dt}{t}=\int\limits_{0}^{T}e^{-t}\sin (\pi \frac{t%
}{T})f(t)dt,\qquad T=\frac{\pi }{2\beta },  \nonumber
\end{equation}%
\begin{equation}
f(t)=\frac{\coth t-\frac{1}{t}}{t}-e^{-T}\frac{\coth (t+T)-\frac{1}{t+T}}{t+T%
}+e^{-2T}\frac{\coth (t+2T)-\frac{1}{t+2T}}{t+2T}-...  \nonumber
\end{equation}%
The integral $I(\beta )$ can be conveniently evaluated numerically or even
analytically \cite{Selg2003}:%
\begin{equation}
I(\beta )=\sum_{n=1}^{\infty }I_{n},I_{n}=\frac{(-1)^{n-1}2^{2n}B_{n}}{%
(2n)(2n-1)(1+4\beta ^{2})^{2n-1}}\sum_{k=0}^{n-1}(-1)^{k}\left( 
\begin{array}{c}
2n-1 \\ 
2k+1%
\end{array}%
\right) (2\beta )^{2k+1}.  \label{F141}
\end{equation}%
Here $B_{n}$ denotes the $n$-th order Bernoulli number ($B_{1}=1/6,$ $%
B_{2}=1/30,$ etc.).

Thus, returning to the main subject of this section, we have found another
option for calculating the phase shift, based on Eqs. (\ref{F13}), (\ref%
{F100}), (\ref{F131}), (\ref{F134}) and (\ref{F138})-(\ref{F141}):

\begin{equation}
\delta (k)=\arctan \left[ \frac{F_{0}\alpha _{N}^{2}\beta \exp
(y_{0})y_{0}^{-(2N+1)}-\tan (\alpha _{0})}{1-\tan \left( \alpha _{0}\right) }%
\right] ,\ F_{0}\equiv F\left( N,i\beta ;y_{0}\right) .  \label{F142}
\end{equation}

This little excursion once again demonstrates wonderful analytic properties
of the Morse potential, providing an instructive example how purely physical
considerations may lead to a pure mathematical result.

\section{Scattering part of the kernel}

As the necessary spectral data are now available, we can start solution of
the inverse problem. First step is to perform the Fourier transform to fix
the scattering part of the kernel (\ref{F3}):%
\begin{equation}
A_{s}(x)\equiv \frac{1}{2\pi }\int\limits_{-\infty }^{\infty }\left[ S(k)-1%
\right] \exp (ikx)dk=-\frac{f(x)+g(x)}{\pi },  \label{F14}
\end{equation}%
\begin{equation}
f(x)\equiv \int\limits_{0}^{\infty }\sin \left[ 2\delta (k)\right] \sin
(kx)dk,\ g(x)\equiv 2\int\limits_{0}^{\infty }\sin ^{2}\left[ \delta (k)%
\right] \cos (kx)dk.  \label{F15}
\end{equation}
\begin{figure}[b]
\includegraphics[width=0.8\textwidth]{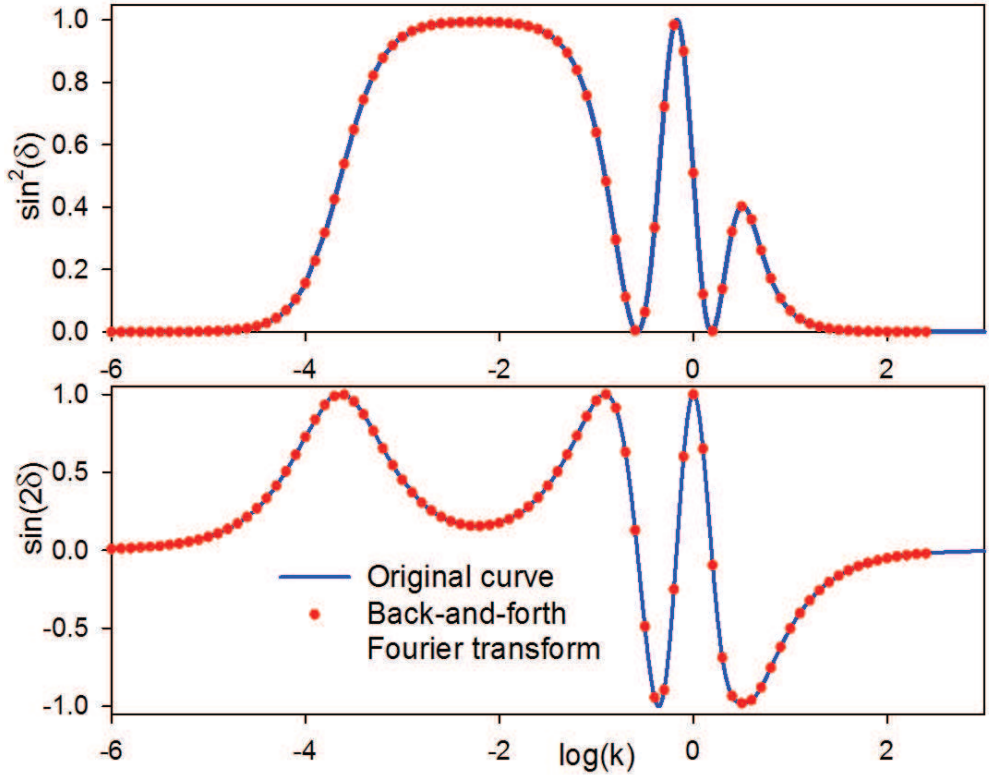} 
\caption{\label{fig:Fourier}
Demonstration of the excellent accuracy of the Fourier transforms performed
according to Eqs. (\ref{F14})-(\ref{F16}).}
\end{figure}

It can be shown that $A_{s}(-x)=\left[ f(x)-g(x)\right] /\pi ,$ so that the
inverse Fourier transform would result in the following formulas:%
\begin{equation}
\sin ^{2}\left[ \delta (k)\right] =\frac{1}{\pi }\int\limits_{0}^{\infty
}g(x)\cos (kx)dx,\ \sin \left[ 2\delta (k)\right] =\frac{2}{\pi }%
\int\limits_{0}^{\infty }f(x)\sin (kx)dx.  \label{F16}
\end{equation}
\begin{figure}[tbh]
\includegraphics[width=0.7\textwidth]{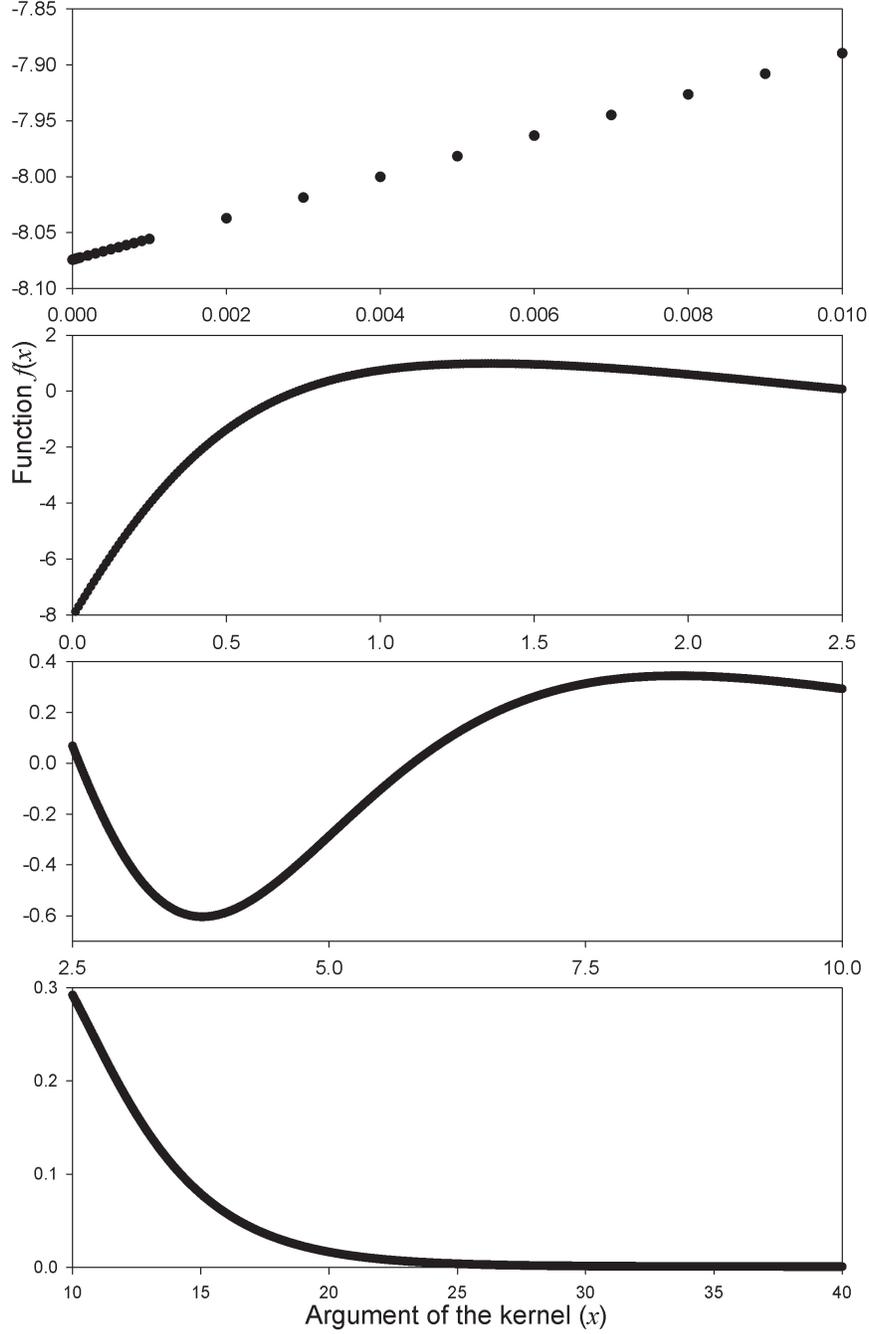} 
\caption{\label{fig:F}Coordinate dependence of the function $f(x)$ calculated
according to Eq. (\ref{F15}).}
\end{figure}
\begin{figure}[tbh]
\includegraphics[width=0.7\textwidth]{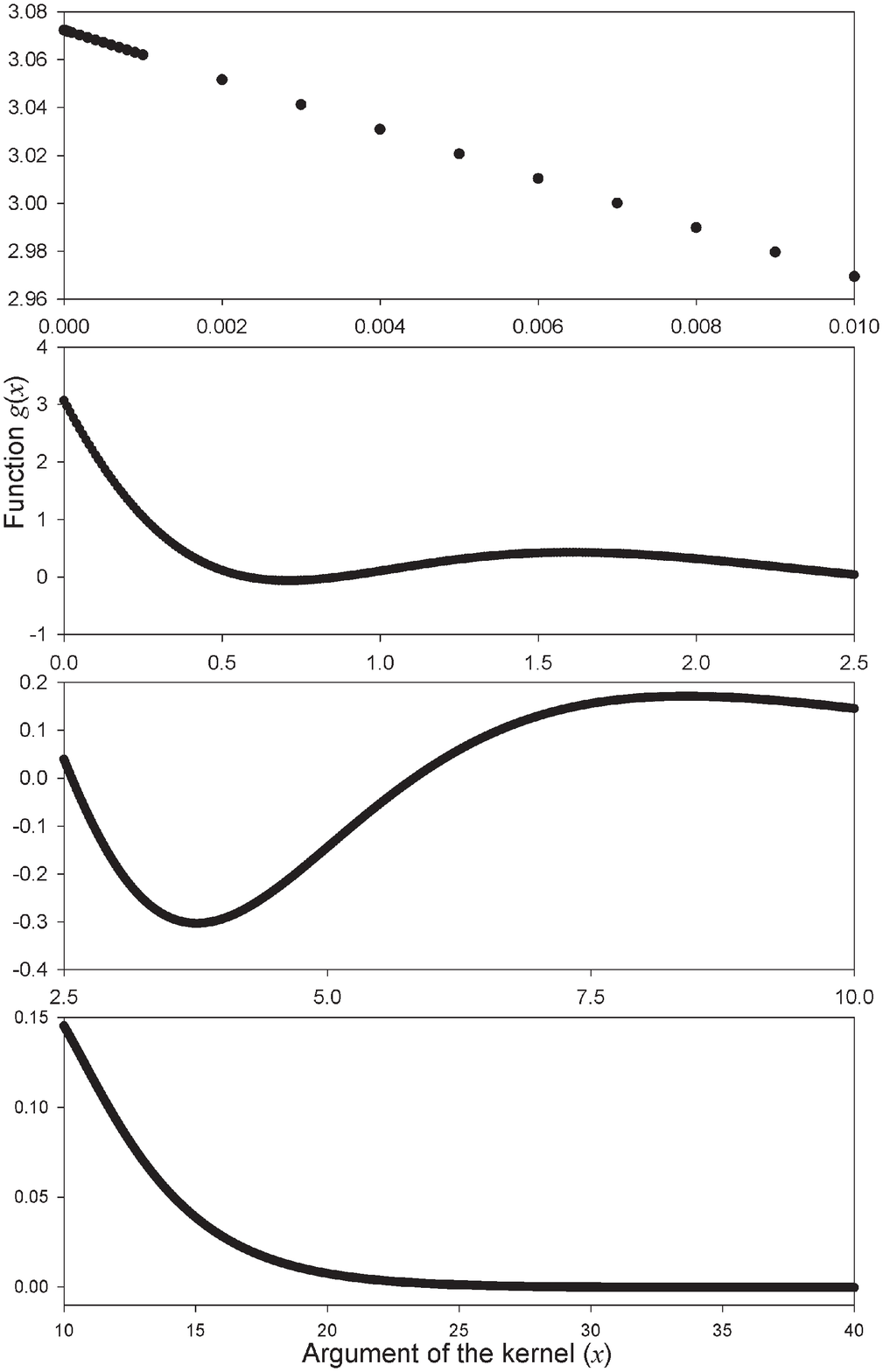} 
\caption{\label{fig:G}Coordinate dependence of the function $g(x)$ calculated
according to Eq. (\ref{F15}).}
\end{figure}
\begin{figure}[tbh]
\includegraphics[width=0.7\textwidth]{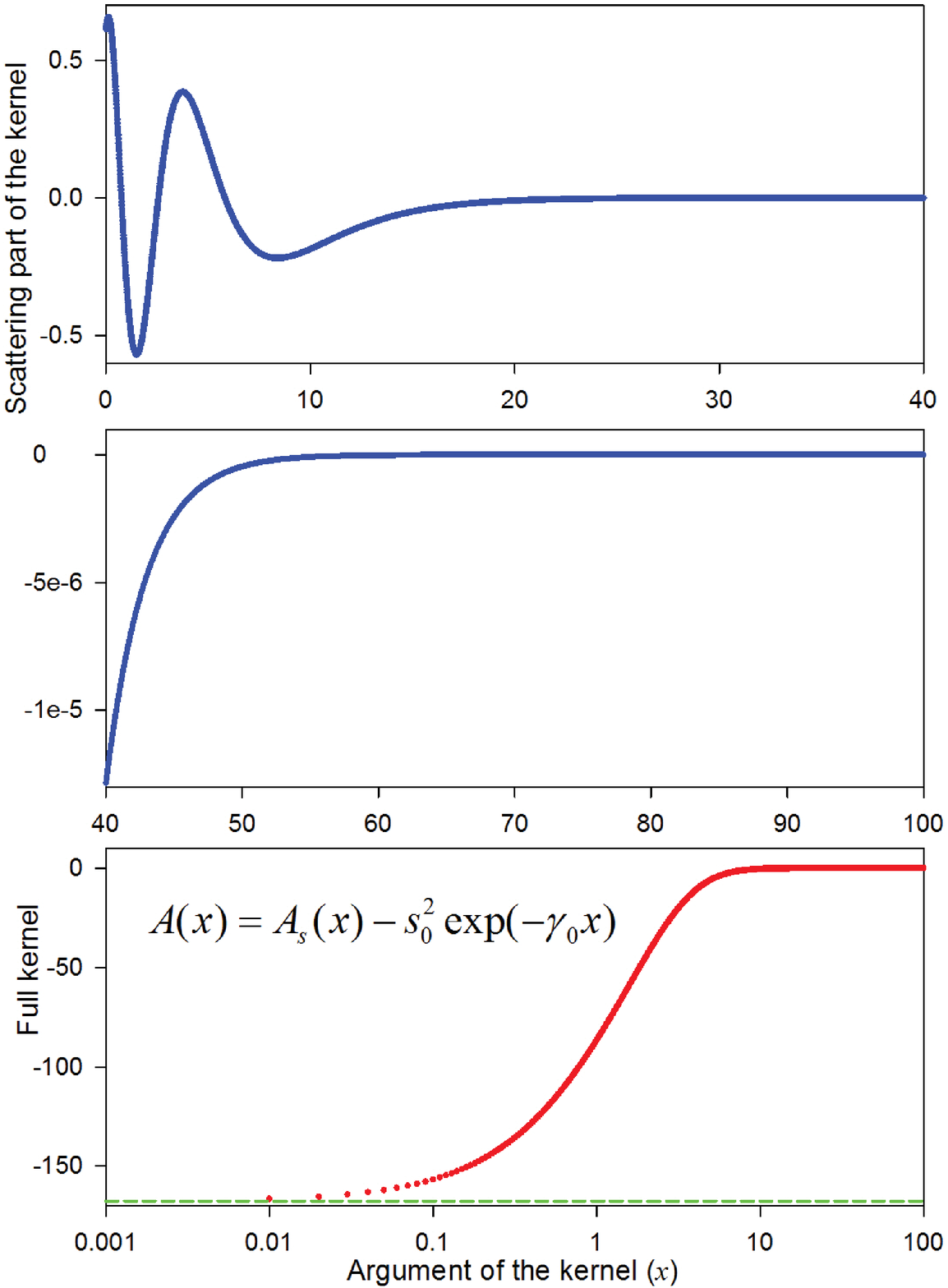} 
\caption{\label{fig:Kernel}Coordinate dependence of the scattering part of the
kernel according to Eq. (\ref{F3}) (two upper graphs) and the full kernel for
the examined Marchenko equation (bottom graph).}
\end{figure}

Calculations according to Eqs. (\ref{F14})-(\ref{F15}) represent the most
challenging part of the overall procedure, because the functions $f(x)$ and $%
g(x)$ must be ascertained in a wide range with an extreme accuracy. It
means, in particular, that the fast Fourier transform techniques are useless
for our purposes. Fortunately, one can use an asymptotic formula \cite%
{Selg2006}%
\begin{equation}
\delta (k)=\frac{a_{1}}{k}+\frac{a_{3}}{k^{3}}+\frac{a_{5}}{k^{5}}%
+...,\qquad k\rightarrow \infty ,  \label{F17}
\end{equation}%
where the coefficients $a_{1},a_{3},a_{5},...$ are directly related to the
potential and its derivatives (for example, $a_{1}=-(2C)^{-1}\int%
\limits_{0}^{\infty }V(r)dr$). Thus the asymptotic part of the integrals (%
\ref{F15}) can be calculated analytically. In addition, one can make use of
the Filon's quadrature formulas (see Ref.~\onlinecite{AbrSteg}, p. 890), which
are perfectly suitable if $x$ is large. The Filon's formulas have been used
for the range $k\in \left[ 20,100\right] $ when $x\geq 20,$ while for $x<20$
a more common (but very accurate) approach was used. Namely, the domain was
divided into sufficiently small intervals where 64-point Gauss-Legendre
quadrature formula is appropriate to the purpose.

The described procedure ensures the correct asymptotic behavior of both
functions defined in Eq. (\ref{F15}). Namely,%
\begin{equation}
f(x)\approx \frac{\pi }{4}y_{0}\exp (-\frac{\alpha x}{2})+b\exp (-cx),\
g(x)\approx \frac{\pi }{4}y_{0}\exp (-\frac{\alpha x}{2})-b\exp (-cx),
\label{F18}
\end{equation}%
where $b=7.252681534782$e-4, $c=2.315574387346$e-4 are some characteristic
constants. Quite remarkably, these formulas hold to better than 10$^{-10}$
accuracy for $x\geq 30.$ The second terms in Eq. (\ref{F18}) are not
important for the following analysis, since they mutually compensate each
other. On the other hand, without these additional terms one cannot
correctly perform the inverse Fourier transform according to Eq. (\ref{F16}%
). In fact, there is no need for this inverse transform as well, but it is
recommended, to be sure that the calculated functions $f(x)$ and $g(x)$ are
reliable. Fig. \ref{fig:Fourier} demonstrates that this is indeed the case.

Figures \ref{fig:F}, \ref{fig:G} and \ref{fig:Kernel} give more detailed
information about the solution procedure. (Although the curves in these
figures may look like solid lines, they are assemblies of dots which
represent the actual results of calculations). Looking at Figs. \ref{fig:F}-%
\ref{fig:Kernel}, one can understand why the utmost accuracy is needed in
performing the Fourier transform. This is most clearly seen in the lowest
graph of Fig. \ref{fig:Kernel} (the green dashed line there shows $A(0)$).
Indeed, if the Fourier transform is not sufficiently correct then the
important information about the scattering properties of the system (shown
in figures \ref{fig:F}, \ref{fig:G} and in the two upper graphs of Fig. \ref%
{fig:Kernel}) may easily be lost against the background of the bound states'
contribution to the kernel, which dominates at small values of the argument.
Thus the whole procedure would become meaningless.

As the result of the Fourier transform according to Eq. (\ref{F15}), one
gets two smooth functions that can be very accurately (with absolute error
less than 10$^{-6}$) approximated by piecewise rational functions%
\[
\left\{ 
\begin{array}{c}
f(x) \\ 
g(x)%
\end{array}%
\right. =\frac{a+bx+cx^{2}+dx^{3}+ex^{4}+fx^{5}}{%
1+gx+hx^{2}+ix^{3}+jx^{4}+kx^{5}+lx^{6}}, 
\]%
whose parameters are given in Tables \ref{tab:table1} and \ref{tab:table2}.
Using these parameters and applying Eq. (\ref{F18}) for larger arguments,
one can reproduce all curves shown in Figs. \ref{fig:F}-\ref{fig:Kernel}.

\begin{table*}[tbp]
\caption{The fitting parameters for the curves shown in
Fig. \protect\ref{fig:F}.}
\label{tab:table1}%
\begin{tabular}{c|cccc}
\hline\hline
& \multicolumn{4}{|c}{Parameters of the rational fit for the function $f(x)$}
\\ 
& $0\leq x\leq 0.2$ & $0.2<x\leq 2.5$ & $2.5<x\leq 10$ & $10<x\leq 30$ \\ 
\hline
$a$ & \multicolumn{1}{|l}{-8.0742834752365} & \multicolumn{1}{l}{
-8.07451927770804} & \multicolumn{1}{l}{0.710177479533} & \multicolumn{1}{l}{
-1.318246783662} \\ 
$b$ & \multicolumn{1}{|l}{-139.70663281} & \multicolumn{1}{l}{21.09396832161}
& \multicolumn{1}{l}{0.078971528944} & \multicolumn{1}{l}{0.350978767195} \\ 
$c$ & \multicolumn{1}{|l}{-1964.01791531} & \multicolumn{1}{l}{
-20.103272203664} & \multicolumn{1}{l}{-0.276506389451} & \multicolumn{1}{l}{
-0.025997992558} \\ 
$d$ & \multicolumn{1}{|l}{9295.18176575} & \multicolumn{1}{l}{10.475159850219
} & \multicolumn{1}{l}{0.065047583073} & \multicolumn{1}{l}{8.813724334413e-4
} \\ 
$e$ & \multicolumn{1}{|l}{-8951.34532192} & \multicolumn{1}{l}{
-2.595277468302} & \multicolumn{1}{l}{-4.695098041353e-3} & 
\multicolumn{1}{l}{-1.492795622311e-5} \\ 
$f$ & 0 & \multicolumn{1}{l}{0.196451244097} & \multicolumn{1}{l}{
1.138463611195e-4} & \multicolumn{1}{l}{1.117794694102e-7} \\ 
$g$ & \multicolumn{1}{|l}{19.59774821} & \multicolumn{1}{l}{-0.316168857071}
& \multicolumn{1}{l}{-0.711060782617} & \multicolumn{1}{l}{-0.345296699762}
\\ 
$h$ & \multicolumn{1}{|l}{287.27716444} & \multicolumn{1}{l}{0.804977555407}
& \multicolumn{1}{l}{0.277492349329} & \multicolumn{1}{l}{0.068087218353} \\ 
$i$ & \multicolumn{1}{|l}{-511.76899191} & \multicolumn{1}{l}{-0.412291982417
} & \multicolumn{1}{l}{-0.053313022152} & \multicolumn{1}{l}{
-6.848766446371e-3} \\ 
$j$ & \multicolumn{1}{|l}{-357.91228265} & \multicolumn{1}{l}{0.355627589898}
& \multicolumn{1}{l}{6.620633459645e-3} & \multicolumn{1}{l}{
4.849331544015e-4} \\ 
$k$ & \multicolumn{1}{|l}{-747.01469965} & \multicolumn{1}{l}{-0.117132459825
} & \multicolumn{1}{l}{-3.875251671984e-4} & \multicolumn{1}{l}{
-1.655209945751e-5} \\ 
$l$ & 0 & \multicolumn{1}{l}{0.024943017758} & \multicolumn{1}{l}{
1.561945583680e-5} & \multicolumn{1}{l}{4.455108455684e-7} \\ \hline\hline
\end{tabular}%
\end{table*}

\begin{table*}[tbp]
\caption{The fitting parameters for the curves shown in
Fig. \protect\ref{fig:G}.}
\label{tab:table2}%
\begin{tabular}{c|cccc}
\hline\hline
& \multicolumn{4}{|c}{Parameters of the rational fit for the function $g(x)$}
\\ 
& $0\leq x\leq 0.25$ & $0.25<x\leq 2.42$ & $2.42<x\leq 9.2$ & $9.2<x\leq 42$
\\ \hline
$a$ & \multicolumn{1}{|l}{3.072175507081882} & \multicolumn{1}{l}{
3.070830178732} & \multicolumn{1}{l}{-4.755191843506e-5} & 
\multicolumn{1}{l}{-0.350846307261} \\ 
$b$ & \multicolumn{1}{|l}{19665382.109603} & \multicolumn{1}{l}{
-0.08759364482} & \multicolumn{1}{l}{0.149350091364} & \multicolumn{1}{l}{
0.087431813456} \\ 
$c$ & \multicolumn{1}{|l}{-57958653.686259} & \multicolumn{1}{l}{
-24.726764443843} & \multicolumn{1}{l}{-0.015988794423} & \multicolumn{1}{l}{
-5.295698261860e-3} \\ 
$d$ & \multicolumn{1}{|l}{37126643.692319} & \multicolumn{1}{l}{
34.416513599217} & \multicolumn{1}{l}{-0.032070876143} & \multicolumn{1}{l}{
1.228492194728e-4} \\ 
$e$ & \multicolumn{1}{|l}{14117338.25322} & \multicolumn{1}{l}{
-13.367864496127} & \multicolumn{1}{l}{6.826423264373e-3} & 
\multicolumn{1}{l}{-7.497417661209e-7} \\ 
$f$ & \multicolumn{1}{|l}{-11218702.142292} & \multicolumn{1}{l}{
1.423672500917} & \multicolumn{1}{l}{-2.754362667700e-4} & 
\multicolumn{1}{l}{-9.379725943814e-9} \\ 
$g$ & \multicolumn{1}{|l}{6400798.980337} & \multicolumn{1}{l}{3.338766375967
} & \multicolumn{1}{l}{-1.199750111544} & \multicolumn{1}{l}{-0.403835495838}
\\ 
$h$ & \multicolumn{1}{|l}{2752650.386055} & \multicolumn{1}{l}{0.570620534536
} & \multicolumn{1}{l}{0.712576159696} & \multicolumn{1}{l}{0.082504174267}
\\ 
$i$ & \multicolumn{1}{|l}{3804036.247907} & \multicolumn{1}{l}{1.953451792127
} & \multicolumn{1}{l}{-0.216463815475} & \multicolumn{1}{l}{
-8.976469204382e-3} \\ 
$j$ & \multicolumn{1}{|l}{-49419.924115} & \multicolumn{1}{l}{-0.164502127028
} & \multicolumn{1}{l}{0.03998655984} & \multicolumn{1}{l}{6.288995558866e-4}
\\ 
$k$ & \multicolumn{1}{|l}{-268374.263668} & \multicolumn{1}{l}{0.34471652522}
& \multicolumn{1}{l}{-3.895572554791e-3} & \multicolumn{1}{l}{
-2.318297714188e-5} \\ 
$l$ & \multicolumn{1}{|l}{1638562.625682} & \multicolumn{1}{l}{0.011395750684
} & \multicolumn{1}{l}{2.046893098608e-4} & \multicolumn{1}{l}{
5.388478464471e-7} \\ \hline\hline
\end{tabular}%
\end{table*}

\section{Solution of Marchenko equation}

Having completed the preliminary work, there remains the last step - the
actual solution of the Marchenko equation (\ref{F2}). To some surprise, this
is technically much easier than determining the kernel (\ref{F3}). For any
fixed $r,$ let us define $T(x)\equiv A(r,r+x),$ $t\equiv r+x$, and $s\equiv
r+y.$ Eq. (\ref{F2}) can then be rewritten as%
\begin{equation}
T(x)=A_{0}(2r+x)+\int\limits_{0}^{\infty }A_{0}(2r+x+y)T(y)dy\approx
A_{0}(2r+x)+\sum\limits_{i}W_{i}F(X_{i}),  \label{F19}
\end{equation}%
where $F(y)\equiv A_{0}\left( 2r+x+y\right) T\left( y\right) ,$ while $X_{i}$
and $W_{i}$ are the nodes and weights of an appropriate quadrature formula,
respectively.
\begin{figure}[tbh]
\includegraphics[width=0.9\textwidth]{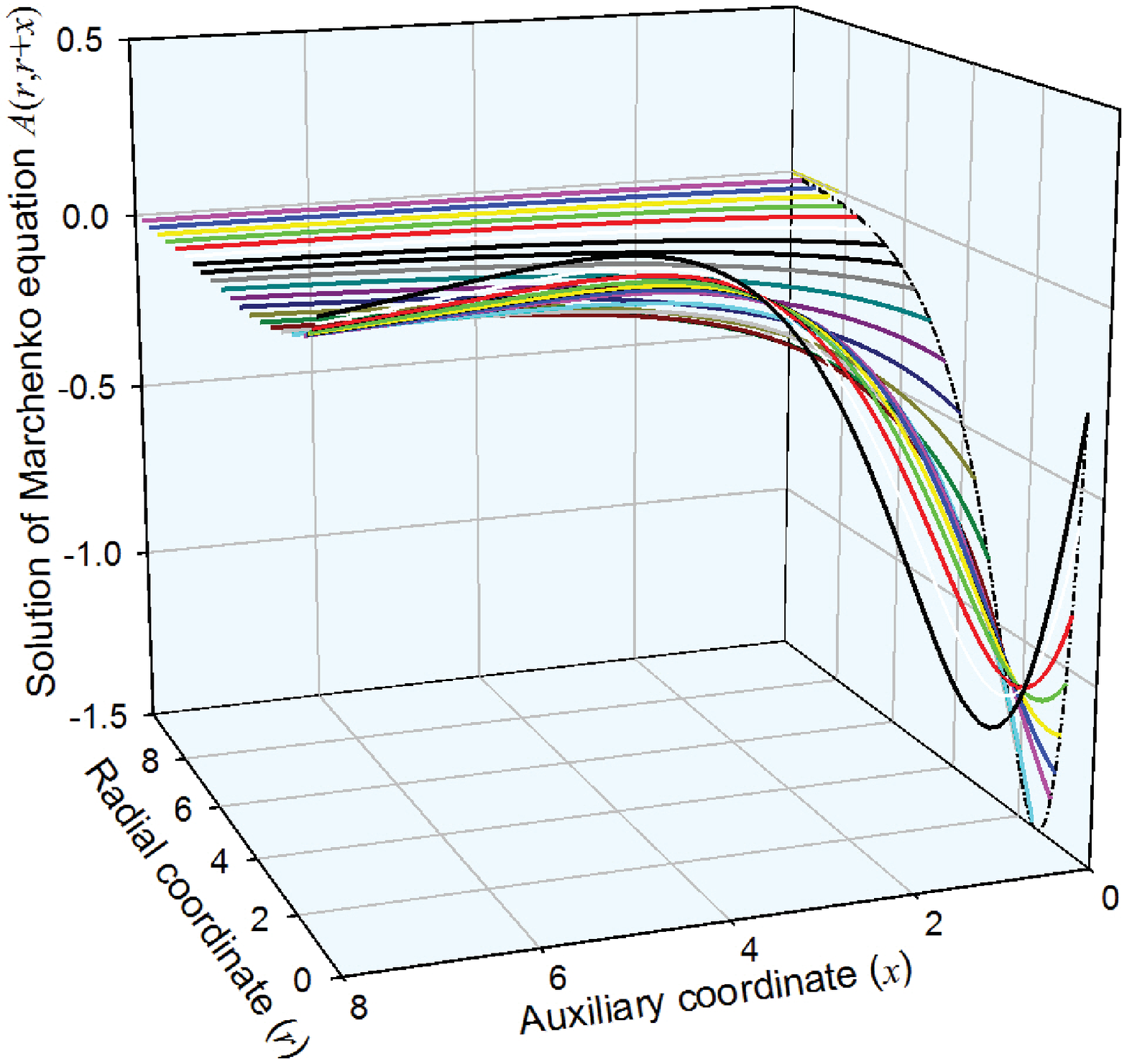} 
\caption{\label{fig:Solution}
3D depiction of a solution to Eq. (\ref{F2}) (see the explanations in the
text).}
\end{figure}
\begin{figure}[tbh]
\includegraphics[width=0.75\textwidth]{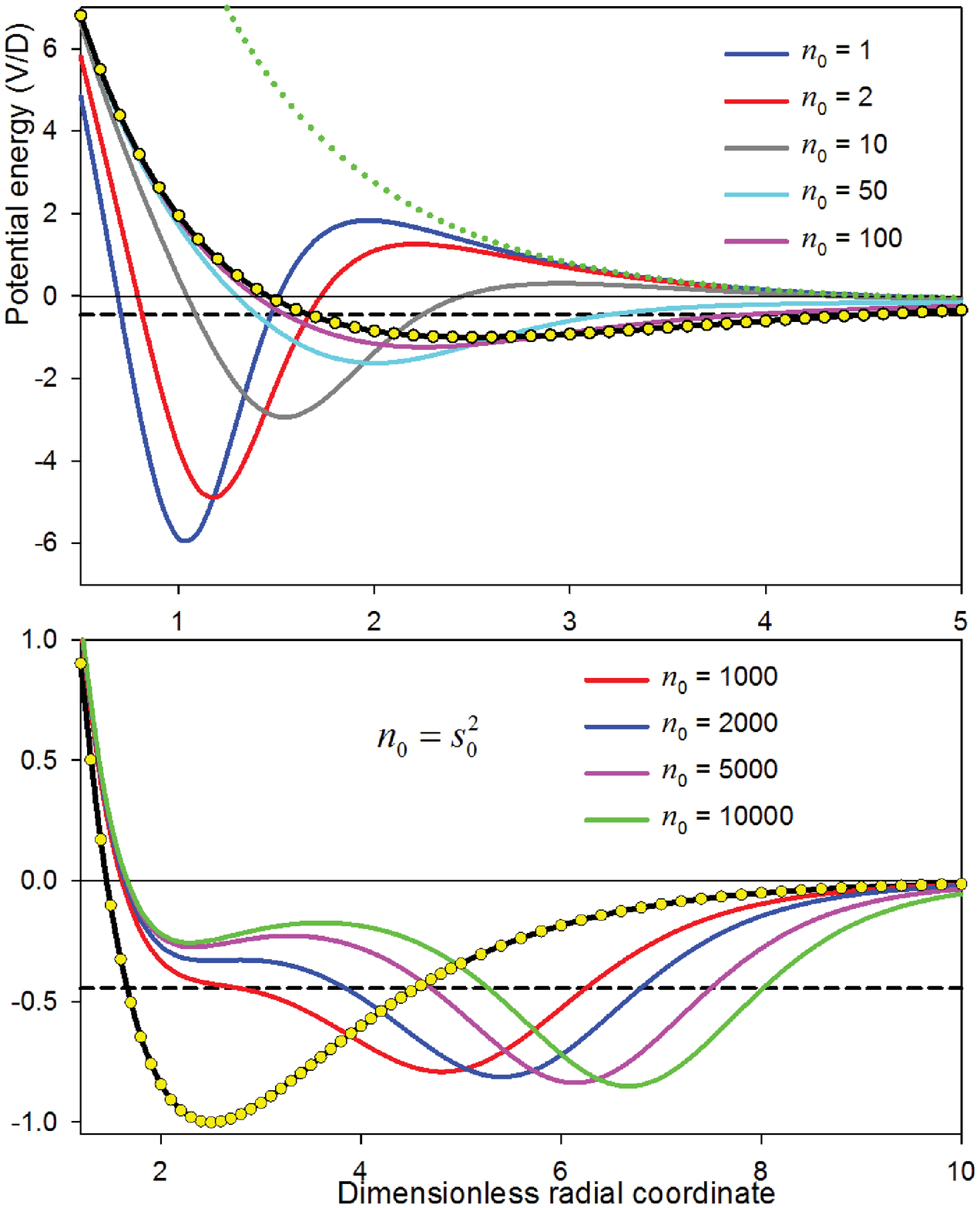} 
\caption{\label{fig:Pots}
A family of phase-equivalent isospectral potentials calculated according to
Eqs. (\ref{F2})-(\ref{F6}). The black solid line shows the initial Morse
potential, yellow circles correspond to the exact theoretical norming
constant $s_{0}^{2}=y_{0}^{2}\alpha =168.18975,$ black dashed line indicates
the position of the energy level $E_{0}=-\gamma _{0}^{2}=4/9,$ and green
dotted line demonstrates the result of 'removing' this level ($s_{0}=0$).}
\end{figure}

Eq. (\ref{F19}) has been solved using the Nystr\"{o}m method (see, e.g., 
Ref.~\onlinecite{NumRec}, p. 782), which is perfectly suitable for our purpo%
ses. As the asymptotic behavior of the kernel is known, a good idea is to split
the integral into two parts and discretize them differently. Namely, for $%
I_{1}\equiv \int\limits_{0}^{R}F(y)dy$ the 64-point Gauss-Legendre
quadrature formula can be successfully used. More specifically, the family
of isospectral potentials in Fig. \ref{fig:Pots} has been calculated with $%
R=15,$ and splitting the range $\left[ 0,R\right] $ into two subdomains, so
that there are 128 nodes within this region.

To discretize the asymptotic integral $I_{2}\equiv \int\limits_{R}^{\infty
}F(y)dy,$ a useful trick is to change the variable \cite{Delves}: $%
y=R+\alpha _{0}(1-u)/(1+u),$ where $\alpha _{0}>0$ is an arbitrary constant.
Correspondingly, the domain of integration reduces to $\left[ -1,1\right] ,$
where one can again use the 64-point Gauss-Legendre quadrature formula. Thus%
\begin{equation}
I_{2}=2\alpha _{0}\int\limits_{-1}^{1}\frac{F(u)}{\left( u+1\right) ^{2}}%
du=\sum\limits_{i}w_{i}^{\prime }F(x_{i}^{\prime }),  \label{F20}
\end{equation}%
where $x_{i}^{\prime }\equiv R+\alpha _{0}(1-x_{i})/(1+x_{i}),$ $%
w_{i}^{\prime }\equiv 2\alpha _{0}w_{i}/\left( x_{i}+1\right) ^{2}$, and $%
x_{i},w_{i}$ are the nodes and weights of the Gauss-Legendre formula,
respectively.

The computational procedure itself is simple. Namely, using the discretized
version of Eq. (\ref{F19}) at the specified node points $x_{n}$, one gets a
system of linear algebraic equations for the quantities $T_{n}\equiv
T(x_{n}) $. Such a system can be easily solved. In this paper, the
Householder reduction algorithm \cite{Householder} was used for this
purpose, because it is numerically more stable than Gaussian elimination.
Having found the solution, one uses Eq. (\ref{F19}) once again, taking
$x=0$ to get $T(0)=A(r,r).$ Finally, one calculates the potential
according to Eq. (\ref{F6}).

The results of solving the Marchenko equation are shown in Figs. \ref%
{fig:Solution} and \ref{fig:Pots}. The curves there were calculated with $%
\alpha _{0}=\Delta \cdot (1+x_{0})/(1-x_{0})$ and $\Delta =10000$ (this
parameter characterizes the width of the domain). As can be seen,
discretization of Marchenko equation with only 192 nodes related to the
relevant quadrature formulas is sufficient to accurately solve the inverse
problem. In addition, one can convince himself/herself that the sign of the
second term in Eq. (\ref{F3}) is indeed 'minus', while $s_{0}^{2}=y_{0}^{2}%
\alpha =6\exp (10/3)$, as expected.

Fig. \ref{fig:Solution} demonstrates the dependence of the solution $%
A(r,r+x) $ on both arguments, $r$ and $x$. In this case, the exact
theoretical value for the norming constant ($s_{0}^{2}=y_{0}^{2}\alpha $)
was used to fix the kernel. Actually, to calculate the potential according
to Eq. (\ref{F6}), one only needs the solution at $x=0,$ which is shown as a
background contour (dash-dot line).

In Fig. \ref{fig:Pots}, a family of isospectral potentials can be seen. It
was obtained by varying the norming constant $s_{0}$ in a wide range (from 0
to 100), while the scattering part of the kernel remains exactly the same
for all these curves. Looking at Fig. \ref{fig:Pots}, one can imagine that
if the real norming constant is not known, a criterion of 'reasonableness'
could be used to fix it. For example, assuming that the potential has only
one extremum point (minimum) and only one inflection point (as the Morse
potential), the correct norming constant can be determined with reasonable
accuracy.

\section{Conclusion}

In this paper, a detailed description of the full solution procedure of the
Marchenko integral equation was given. To this end, the complete set of the
necessary input data was used, which was obtained by accurately solving the
direct Schr\"{o}dinger problem for a model system. Such a tautological
combination of the direct and inverse scattering problems may seem useless
for practical purposes, but one has to bear in mind that otherwise the full
set of necessary data cannot be obtained at all. Unfortunately, the methods
of the inverse scattering theory are still not implemented as planned. These
methods have mainly been used to construct families of isospectral
potentials \cite{AM, Pursey}, and they have many interesting applications in
the soliton theory (see, e.g., \cite{EvH, Ablowitz}), but their capabilities
have not been fully exploited. Hopefully, the results of this work clearly
demonstrate the real power of the Marchenko method, albeit the necessary
input data were obtained computationally, not experimentally. On the other
hand, the model system used for this purpose, the well-known Morse
potential, continues to surprise us by its undiscovered analytical
properties. For example, in this paper a simple formula for calculating the
phase shift was derived, along with the new analytic algorithm for
calculating the Riemann-Siegel function, which is a pure mathematical result.

How could the results of this work be useful for further studies? An
interesting option is to combine Eq. (\ref{F2}) with the Marchenko
differential equation (see, e.g., Ref.~\onlinecite{Chadan}, p. 78), assuming
that the real potential can be expressed as a sum $V(r)=V_{0}(r)+\Delta V$,
where $V_{0}(r)$ is a known model potential whose spectral characteristics are
known as well. Thus according to Eq. (\ref{F6}), $V_{0}(r)=-2\left[
A_{0}(r,r)\right] ^{\prime },$ where $A_{0}(r,t)$ is the corresponding
solution to Eq. (\ref{F2}). Then, assuming that $V(r)=-2\left[ A(r,r)\right]
^{\prime }$ and $A(r,t)=A_{0}(r,t)+\Delta A$, one gets an equation%
\begin{equation}
\left( \frac{\partial ^{2}}{\partial r^{2}}-\frac{\partial ^{2}}{\partial
t^{2}}\right) \Delta A(r,t)=V_{0}(r)\Delta A(r,t)+\Delta V(r)A_{0}(r,t).
\label{F21}
\end{equation}

If $V_{0}(r)$ is sufficiently close to the real potential, the right side of
Eq. (\ref{F21}) would contain two small quantities, $\Delta V$ and $\Delta A$%
, which, in principle, can be determined self-consistently along with
solving this equation.

\section*{Acknowledgements}

The author acknowledges support from the Estonian Ministry of Education and
Research (target-financed theme IUT2-25) and from ERDF (project
3.2.1101.12-0027) for the research described in this paper.

\end{document}